**Advances in aviation radiation mitigation were demonstrated during the Gannon storm**

W. K. Tobiska, B. Hogan, L. Didkovsky, K. Judge, J. Bailey, K. Drumm, K. Wahl, and A. Sosnov
Space Environment Technologies, Pacific Palisades, CA 90272 USA

# Abstract

The validation of a strategy for aviation radiation hazard mitigation in development for decades has been completed using two commercial airline flights in 2024 and 2025. This article provides a historical review of the primary elements leading to that strategy, including the emergence of aviation radiation awareness and collaborative efforts by global aviation and radiological bodies that established mitigation standards. The primary radiation sources, galactic cosmic rays (GCRs) and solar energetic particles (SEPs), and their mechanisms of impact on Earth's atmosphere are summarized. The article highlights the biological effects of radiation exposure influenced by altitude, latitude, and geomagnetic conditions, upon aircrew, frequent flyers, and commercial space travelers. It recognizes a recent *Space Weather Advisory Group* (SWAG) report that identifies the need for continuous monitoring and predictive models to ensure long-term occupational and public health safety. With this backdrop the validation of an *As Low As Reasonably Achievable* (ALARA) strategy was accomplished using two UAL 990 flights on B777-200 aircraft between San Francisco and Paris. Each carried the same ARMAS FM7 radiation monitoring unit, where one flight occurred during the extreme geomagnetic storm (Gannon storm) May 10–11, 2024 and one flight occurred during quiet geomagnetic conditions June 8–9, 2025. The flights' results validated the strategy during extreme space weather, i.e., by applying operational controls for shielding to reduce dose. One approach is flying lower magnetic latitudes to gain more Earth magnetic field shielding and the other is flying lower altitudes to use atmosphere depth shielding. Both ALARA shielding methods are controllable in airline operations and air traffic management. These have now been validated with total dose measurements by ARMAS. This study shows the effectiveness of the strategy to deviate flight paths to lower magnetic latitude routes and lower altitudes during major geomagnetic storms. Not only does this approach mitigate HF communication outages but it also reduces risks from increased GNSS errors for take-off and landing navigation. Magnetic field shielding is a major risk reduction factor for radiation, communication, and navigation while altitude shielding reduces radiation hazard risks.

# Highlights

- Two commercial airline flights San Francisco and Paris carried the same ARMAS FM7 radiation monitoring unit, where one flight occurred during the extreme geomagnetic storm (Gannon storm) May 10–11, 2024 and one flight occurred during quiet geomagnetic conditions June 8–9, 2025.
- The flights' results validated the ALARA strategy that multiple stakeholders in aviation radiation hazard mitigation have been pursuing for decades, i.e., use operational decisions during extreme space weather to apply shielding by flying lower magnetic latitudes to gain more Earth magnetic field shielding and flying lower altitudes to use atmosphere depth shielding.

- A third shielding method provided by nature in the G5 storm is from Forbush decreases at the beginning of a storm that reduce the number of lower energy protons entering the Earth's atmosphere and lower the "floor" of GCR radiation exposure from cosmic rays.

# 1 Introduction

## 1.1. Historical overview of aviation radiation awareness

The awareness of aviation radiation developed alongside advancements in aerospace technology and the study of atmospheric phenomena. Research in atmospheric physics during the early 20$^{th}$ Century, notably the discovery of cosmic rays through balloon experiments [Hess, 1912], laid a scientific foundation for understanding radiation in Earth's upper atmosphere. However, it was not until after World War II, when commercial and military aviation began regularly operating at higher altitudes, above 8 km (~26,000 ft.), that the potential impacts of radiation exposure gained attention.

During the Cold War, the increase in high-altitude reconnaissance flights and early space exploration missions further highlighted an importance for understanding the complex radiation environment in near-Earth space. In the late 20$^{th}$ Century, the establishment of space weather research programs, driven by increased solar monitoring from satellites, dramatically enhanced knowledge about solar activity cycles and their effects on Earth's atmosphere and magnetosphere.

These advances combined to foster the realization that aviation radiation is an occupational and public health issue. The International Civil Aviation Organization (ICAO), the U.S. Federal Aviation Administration (FAA), and European aviation regulators began working with radiation protection bodies such as the International Commission on Radiological Protection (ICRP) and the International Standards Organization (ISO) to develop a scientific basis for understanding the sources and effects as well as mitigate radiation exposure at aviation altitudes through standards and best practices. Their collaboration led to the establishment of safety guidelines, recommendations for exposure limits, and standardized dosimetry methods. Ongoing real-time space weather monitoring plus modeling research have produced predictive capabilities that are improving aviation radiation mitigation.

## 1.2. Sources of radiation at aviation altitudes

The primary radiation sources at aviation altitudes are galactic cosmic rays (GCRs) and solar radiation events (solar energetic particles, SEPs). GCRs, which are mostly protons and lower species ions, originate outside the solar system. The fluxes of GCRs are slowly modulated by the strength of the Sun's interplanetary magnetic field [Simpson, 1983]. SEPs come from solar activity such as coronal mass ejections (CMEs) related to flaring events or from interplanetary shocks [Gopalswamy, 2004; Reames, 2013]. In the latter case, fast CMEs plow through both ambient background and high-speed stream solar wind fields to create a shock front that produces accelerated energetic protons. The interaction of the GCR and SEP energetic particles with the Earth's troposphere and mesosphere produce additional secondary radiation. Once they enter the atmosphere, they impact $O_2$ and $N_2$ to explode those molecules and create secondary high energy particles such as neutrons, protons, and muons. Recently, precipitated charged particles (PCPs) from Van Allen radiation belt (VAB) interactions have been proposed as an additional radiation source [Tobiska *et al.,* 2018]. Further studies show periods of measured dose rate enhancements may be correlated with hiss wave activity [Aryan *et al.,* 2023; 2025]. Beyond hiss waves, other emissions have also been shown to accelerate VAB particles to relativistic energies, including chorus waves [Drozdov et al., 2020] especially during times of a preconditioned magnetosphere [Allen et al., 2023]. The exact mechanism of producing higher radiation from non GCR/SEP sources is still being discovered.

### 1.3. Effects of radiation at aviation altitudes

Radiation exposure occurs when energetic particles and photons impact tissue molecules and DNA, causing sites for pre-cancerous cell activity. This increased risk from impacts by large numbers of energetic particles and photons varies with altitude, latitude, and geomagnetic activity. Higher altitude (> 8 km) and higher latitude air traffic routes above 8 km [Friedberg and Copeland, 2003, 2011; Tobiska *et al.,* 2016] are particularly vulnerable because the Earth's magnetic field offers less shielding in these regions. These factors highlight the importance of monitoring radiation at aviation altitudes to understand the processes, effects, and methods for exposure mitigation for long-term health in aircrew, frequent flyers, early trimester fetuses, and commercial space travelers.

### 1.4. ICRP guidelines, ISO standards, and ICAO SARPS focused on aviation radiation mitigation

Mertens and Tobiska [2021] summarized international, including U.S., regulatory activities related to the exposure from radiation at aviation altitudes. They described both the E.U. and the U.S. activities in this area. Bain et al. [2023] identified further progress that is needed by the aviation community for space weather radiation forecasts. They noted a lack of routine observations for improving radiation modeling from ground-based neutron monitors and airborne radiation measurements, particularly during solar energetic particle events. Beyond these recent overviews, there are three international organizations that have devoted their attention to aviation radiation exposure and its mitigation.

**ICRP.** First, the International Commission on Radiological Protection (ICRP) provides recommendations that specifically address aviation radiation exposure. For example, ICRP Pub. 132 [2016] recognizes that aircrew are occupationally exposed to ionizing radiation at altitude and airlines should manage and/or monitor radiation doses for crew members with occupational exposure classifications, dose data, operational recommendations for monitoring, education, and pregnancy. ICRP Pub. 103 [2007] recommends an occupational dose limit of 20 mSv/year, averaged over 5 years (100 mSv in 5 years), with no single year exceeding 50 mSv. In this document, it recommends pregnant aircrew should have fetal dose limited to 1 mSv during pregnancy. These documents also outline guidelines for: *i)* monitoring and assessment, i.e., airlines should assess exposure using validated dosimetry models or measurement methods, particularly for flights at higher altitudes or polar routes and perform regular dose assessments for frequent flyers or crew routinely exposed to higher doses; *ii)* education and training, i.e., crew members should receive education on cosmic radiation, its health impacts, and protective measures as well as be provided awareness training so that crew members can make informed decisions about exposure risks.; and *iii)* mitigation and protection, i.e., operators should reduce radiation exposure through flight-route planning, altitude adjustments, and scheduling to avoid high radiation during intense solar events.

**ISO.** Second, the International Standards Organization (ISO) Technical Committee (TC) 85 (Nuclear energy, nuclear technologies, and radiological protection) Sub-Committee (SC) 2 (Radiological Protection) Working Group (WG) 21 (Dosimetry for exposures to cosmic radiation in civilian aircraft) has developed a standard covering the aviation radiation environment. ISO 20785 *"Dosimetry for exposures to cosmic radiation in civilian aircraft"* currently has three parts addressing different aspects of aviation radiation measurement and monitoring. The documents provide guidelines and standardized methods for measuring and monitoring cosmic radiation exposure for aircrew and frequent flyers, ensuring consistent approaches to radiation safety management in aviation. IS 20785 focuses on the cosmic ray background component of aviation radiation, which is the primary exposure source.

ISO 20785-1:2020: *Conceptual basis for measurements* specifies the basis for the determination of ambient dose equivalent for the evaluation of exposure to cosmic radiation in civilian aircraft and for the calibration of instruments used for that purpose. ISO 20785-2:2020: *Characterization of instrument response* specifies methods and procedures for characterizing the responses of devices used for the determination of ambient dose equivalent for the evaluation of exposure to cosmic radiation in civilian aircraft. The methods and procedures are intended to be understood as minimum requirements. ISO 20785-3:2023 *Measurements at aviation altitudes* provides the basis for the measurement of ambient dose equivalent at flight altitudes for the evaluation of the exposures to cosmic radiation in civilian aircraft.

ISO TC 20 (Aircraft and space vehicles) SC 14 (Space systems and operations) WG 4 (Space Environment – Natural and Artificial) is developing a New Work Item to become a standard related to aviation radiation that builds on the body of work of IS 20785 for GCRs. It will also incorporate the effects of SEPs and VAB PCPs on the radiation environment.

**ICAO.** Third, in the pre-COVID era, the International Civil Aviation Organization (ICAO) introduced Standards and Recommended Practices (SARPs) via updates to the ICAO Annex 3 – *Meteorological Service for International Air Navigation*, Chapter 3, Section 3.8. These updates addressed the impact of space weather on aviation and the obligations of regional space weather centers to monitor and provide advisories on space weather phenomena and effects related to aviation. The three main effects on aviation from space weather phenomena are: *i)* high frequency (HF) radio communication loss from solar flares and geomagnetic disturbances affecting the ionosphere; *ii)* navigation inaccuracies, particularly during takeoff and landing, from global navigation satellite system (GNSS) outages due to solar flare and geomagnetic disturbance related scintillation in the ionosphere; and *iii)* increased radiation exposure risk from SEP events that is an additive radiation hazard on top of the ubiquitous GCR background radiation exposure. In November 2019 global space weather information service providers were established: *a)* ACFJ consortium (comprising Australia, Canada, France and Japan); *b)* PECASUS consortium (Pan-European Consortium for Aviation Space Weather User Services comprising Austria, Belgium, Cyprus, Finland, Germany, Italy, Netherlands, Poland and the United Kingdom); *c)* United States (USA); and *d)* CRC consortium (China and Russia). Those providers currently operate on a rotational basis of two weeks each for delivering space weather related advisories to the international aviation community.

The SARP guidelines encourage the development of reliable forecasting tools and the integration of space weather advisories into flight planning and operational decision-making. Airlines and aviation authorities are requested to adopt best practices that include rerouting flights during severe space weather events, adjusting altitude to minimize exposure, and implementing advanced shielding measures for critical aircraft systems.

In terms of radiation, the ICAO Annex 3, Chapter 3 calls out the following activities: *i)* provide information on space weather risks related to aviation; *ii)* ensure monitoring through ground based, airborne and space-based observations to detect space weather conditions having an effect on radiation exposure at flight levels; *iii)* develop regional and global space weather centers to provide radiation increased exposure risk; *iv)* identify intensities of radiation increased exposure risk; *v)* produce advisory messages spelling out the details of radiation increased exposure risk; *vi)* identify flight levels (altitude), longitudes, and latitudes for space weather advisory information; *vii)* provide NOtices To Air Men (NOTAMS) for forecasts of space weather events including the date and time of the event, the flight levels where provided, and portions of airspace which could be affected; and *viii)* update ICAO *Procedures for Air Navigation Services – Air Traffic Management* (PANS-ATM, Doc 4444) to transmit information by air traffic services (ATS) to aircraft concerning space weather activity and, in particular, request descents by aircraft due to radiation exposure from space weather events.

### 1.5. SWAG User Needs Survey findings and recommendations for aviation radiation

In the U.S. the Space Weather Advisory Group (SWAG) was commissioned by the Promoting Research and Observations of Space Weather to Improve the Forecasting of Tomorrow (PROSWIFT) Act [2020]. It is an independent advisory body to U.S. Government agencies to address issues related to space weather hazard mitigation. Their second report, *Results of the First National Survey of User Needs for Space Weather* [2024], summarized the broadest national survey to date across stakeholder communities for understanding the risks and mitigation paths for space weather. Survey participants provided input over the course of 3 years to SWAG, which then summarized their comments into key findings and actionable recommendations aimed at addressing risks associated with space weather phenomena. This SWAG report recommends pathways for policy changes.

The Aviation Sector was one of several key sectors identified in the report and among its core conclusions was the importance of advanced monitoring systems to detect the hazard to aviation from radiation sources. Specifically, Finding 3.2 noted: "There is a lack of measurements, reporting, limits, education, and hazard mitigation pathways for radiation exposure across the aviation industry." Under that Finding several recommendations were made, including Recommendations: <u>3.2.1.</u> "NWS, in collaboration with NASA, NSF, and FAA, should conduct or acquire ionizing radiation measurements at all relevant aviation altitudes and make them available for use by the aviation community. Measurements could be acquired via dosimeter badges on flight personnel, instrumenting individual or fleet/commercial/business aircraft, or purchasing data commercially;" and <u>3.2.3.</u> "FAA, NASA, and NOAA, in coordination with industry and academia, should expand their data reporting and data collection mechanisms to the aviation community to obtain scientific measurements that can validate existing models, such as FAA's Civil Aviation Research Institute (CARI) [Copeland, 2017] and NASA's *Nowcast of Atmospheric Ionizing Radiation for Aerospace Safety* (NAIRAS) [Mertens *et al.,* 2013] models. This will provide the aviation industry a better understanding of the impact on human health from radiation exposure at flight altitudes with assimilative modeling." The importance of the report is that it provided, for the first time, *i)* an overarching characterization of the issues related to aviation radiation exposure from the perspective of the stakeholder community and *ii)* pathways for addressing the mitigation of these issues. Participants in that sector particularly advocated for greater integration of space weather data into flight management protocols, enabling airlines to proactively reroute or adjust flight paths during severe solar events.

The measurement of radiation at aviation altitudes is a high priority and the discussion below identifies recent examples of how those measurements are taking place that have validated, for the first time, a clear mitigation strategy to reduce the aviation radiation exposure hazard. The system we discuss is Space Environment Technologies' (SET) *Atmospheric Radiation Measurement for Aerospace Safety* (ARMAS) system, which has conducted over 1317 flights since 2013 to build one of the most extensive aerospace radiation databases in the world.

### 1.6. Recognized strategy for mitigating aviation radiation using time, altitude, and latitude

A broadly recognized strategy for mitigating aviation radiation is ALARA (As Low As Reasonably Achievable). This is a radiation safety principle for minimizing radiation exposure. It articulates a strategy including: *i)* <u>time</u>, i.e., minimize the duration of exposure to radiation sources; *ii)* distance, i.e., increase the distance between a person and the radiation source since radiation intensity decreases with distance; *iii)* shielding, i.e., use appropriate shielding materials to block or reduce radiation exposure; *iv)* regular monitoring, i.e., regularly monitor radiation levels and individual exposure to ensure that efforts to reduce exposure are effective; and *v)* training, i.e., ensure that individuals working with radiation are adequately trained on safe work practices.

Mitigation of aviation radiation exposure has followed the ALARA principle. For example, lowering flight altitude and latitude during high-radiation events is an effective shielding method to reduce aviation radiation risks. In the altitude domain for middle latitudes, every 2 km lower reduces dose by half using atmospheric shielding; conversely every 2 km higher increases the dose by a factor of 2. Flying at lower altitudes during solar storm events reduces exposure to atmospheric secondary radiation while maintaining operational safety. By using routes that have lower magnetic latitudes also takes advantage of Earth's magnetic field cutoff rigidity shielding. Flights can be rerouted to lower latitudes where Earth's magnetic field offers greater protection from cosmic rays and solar energetic particles. These strategies have now been confirmed for the first time during storm conditions using ARMAS.

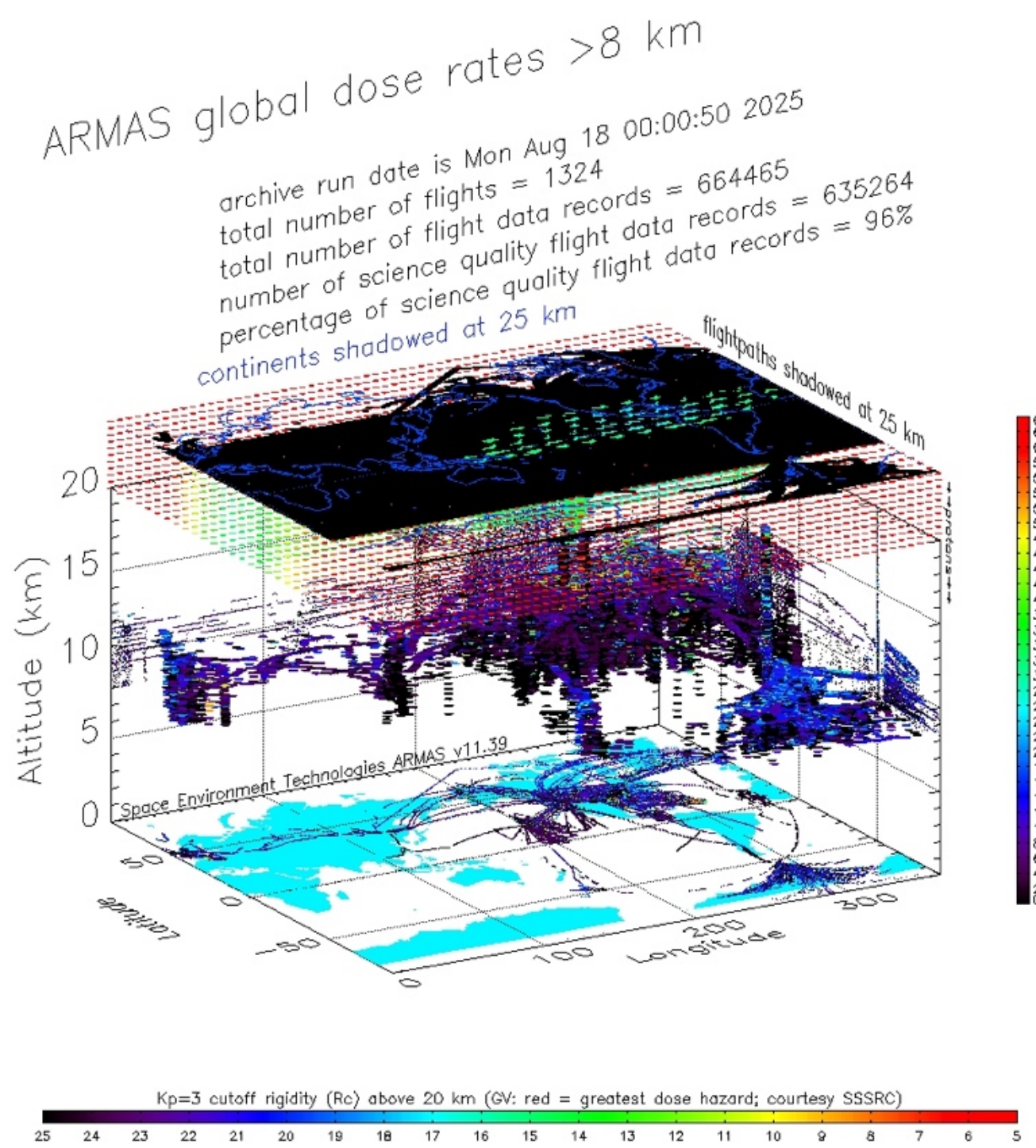

Fig. 1. All ARMAS global measurements are presented from 2013–2025 from the atmosphere above 8 km out to the ISS near 500 km. Each dot represents one of 633910 1-minute science quality data records viewed by latitude, longitude and altitude. The dot color is identified on the Y-axis colorbar. This database contains 1317 flights from all vehicle types (aircraft, suborbital and orbital, balloons) on which ARMAS has flown. Most magnetic latitudes and longitudes are covered. Except for the ISS orbits as the black rectangular region at the top of the plot where 1 day = 1 flight, all flight paths are shown shadowed on the 2D Earth map. The colored dots at the top of the figure indicate cutoff rigidity from Smart and Shea [2009] in values of GV shown by the X-axis colorbar.

# 2 Data and methodology description

## 2.1 ARMAS quantifies aviation radiation

The ARMAS system represents a significant advancement in quantifying radiation exposure during flights from previous measurement methodologies. The ARMAS program utilizes data collection from dozens of aircraft, balloons, suborbital spacecraft, satellites, the International Space Station, and an Earth-Moon flight. It provides quantifiable measurements of cosmic rays, solar energetic particles, and secondary radiation generated in the atmosphere across the planet and at all altitude layers. This system is particularly valuable for immediately assessing radiation levels on

mid- to high-latitude routes, where susceptibility to space weather effects is heightened due to reduced shielding from Earth's magnetic field.

From a database of 1317 flights as of June 2025 (Figure 1), the ARMAS program has used 15 separate instrument types and over 30 unique instruments to accumulate 633910 science quality, 1-minute data records. This database has enabled the generation of the ARMAS statistical model [Tobiska *et al.,* 2018] of the global radiation environment from the surface to 100 km based on altitude, latitude, longitude, and geomagnetic activity. The ARMAS statistical model compares very well with the NAIRAS v3 model and creates a predictive modeling capability that offers airlines and aviation stakeholders actionable information to mitigate risks. ARMAS data, combined with the NAIRAS model into the *RADIation environment using ARMAS data in the NAIRAS model* (RADIAN) data assimilative system, provides optimal flight altitudes and latitudes during high-radiation events, thus ensuring both biological safety for passengers and the operational integrity of onboard electronic systems.

For the data shown below the ARMAS Flight Module 7 (FM7) (Figure 2) system was used to collect radiation dose and dose rate measurements on commercial flights between San Francisco and Paris. We describe the detailed measurements in the next sections. The FM7 is packaged in a rugged housing constructed from milled aluminum and is designed for use on commercial or business jet class aircraft as well as on commercial space travel suborbital vehicles. The FM7 unit does not need to be physically attached to the vehicle. It operates with an external power supply (either commercial-off-the-shelf (COTS) battery or AC aircraft power via a 5 VDC converter). It consists of two components, i.e., the instrument and app.

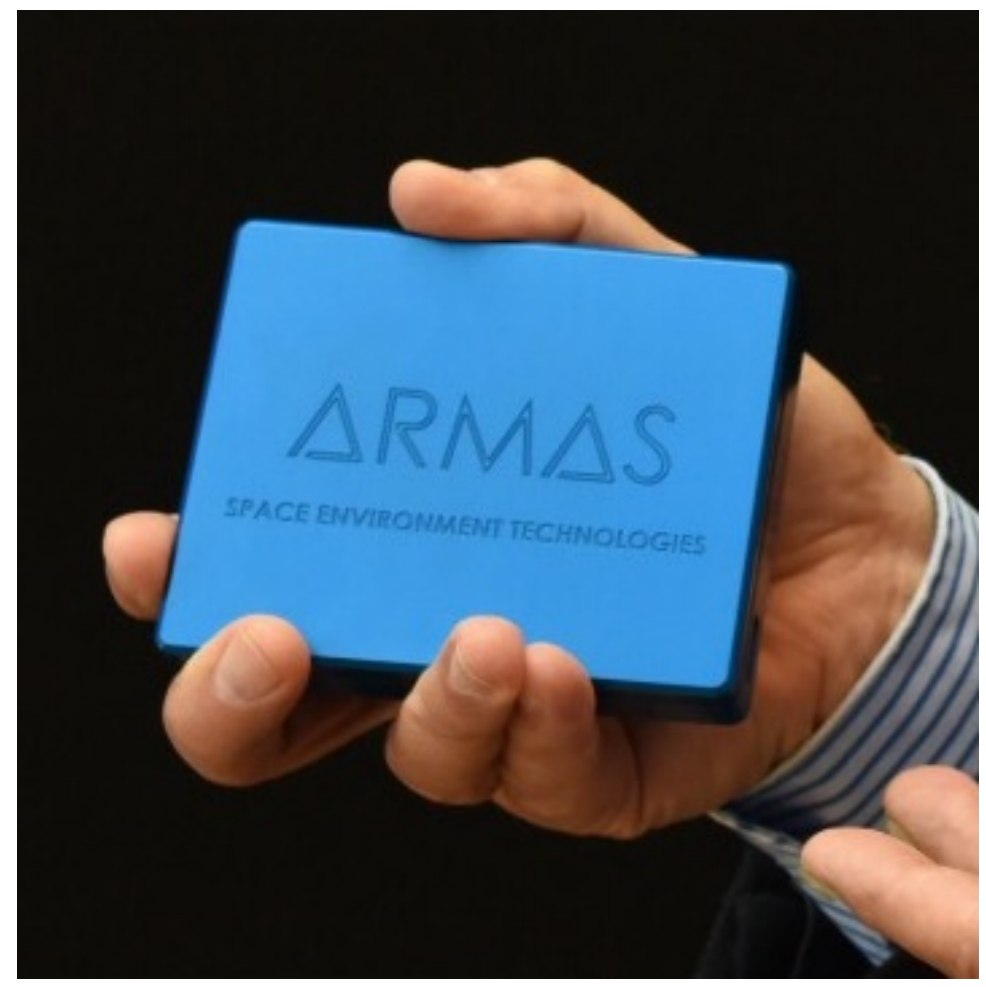

Fig. 2. ARMAS FM7 used in commercial aircraft and suborbital spaceships.

The FM7 flight instrument provides real-time dosimetric measurements of the radiation environment from aircraft or suborbital spacecraft. Radiation dose is created by the penetration of GCRs ($p^+$, α, $Fe^+$), SEPs ($p^+$) and radiation belt particles ($e^-$, $p^+$) into the Earth's atmosphere that subsequently collide with neutral species ($N_2$, $O_2$) to create secondary and tertiary particles and photons (n, $p^+$, $e^-$, α, β, μ, π, γ-rays, X-rays). Measurements are made using a Teledyne micro dosimeter (μDos) UDOS001-C in combination with global positioning system (GPS), Bluetooth, micro-SD data logger, microprocessor, and external power supply. All these are mated to a printed circuit board and provide, via active Bluetooth pairing to an iOS iPhone or iPad app, the real-time absorbed dose rate in silicon, the dose equivalent rate, the ambient dose equivalent rate, and effective dose rate of the radiation environment within the vehicle.

The 10-second real-time data stream is instantaneously available on the ARMAS app via Bluetooth pairing between the app and the FM7 that displays silicon and tissue-relevant dose rates. The real-time data are recorded to a micro-SD data logger whenever FM7 is powered on. Data can also be manually extracted from the micro-SD card after the FM7 is powered off. The real-time data packets can be downlinked to the ground via the app connected to a Wi-Fi network. The downlinked data are processed by ground servers and compared with the NAIRAS global radiation climatology dose data with an approximate 1–minute latency on the ARMAS web site.

## 2.2 The NAIRAS model

An existing state-of-the-art radiation environment model is NASA LaRC's NAIRAS model that produces data-driven, physics-based climatology of time-averaged global radiation conditions [Mertens *et al.,* 2013]. It predicts dosimetric and radiative flux quantities for evaluating radiation exposure levels for humans and electronic systems on flights. There are several features that make NAIRAS stand out. First, it covers the entire domain of interest from the surface of the Earth to deep space using physics-based modeling. It predicts dosimetric quantities from the surface of the Earth, through the atmosphere, and into LEO from GCR, SEP, trapped proton, and trapped electron sources. Effects of GCRs and SEPs are modeled for species up to uranium (Z = 92, A = 238). NAIRAS considers the local geomagnetic cutoff rigidity and includes the response of the geomagnetic field to interplanetary dynamical processes and subsequent influences on atmospheric dose. It uses coupled physics-based models to transport cosmic radiation through three distinct domains: the heliosphere, Earth's magnetosphere, and neutral atmosphere. Second, the physics-based models are driven by *real-time measurements* to specify boundary conditions on the cosmic and solar radiation at the interfaces between the distinct domains or to characterize domain internal properties through which radiation propagates. Third, NAIRAS is at *Technology Readiness Level (TRL) 9*, i.e., fully integrated with operational hardware and software systems, user documentation completed, all functionality tested in operational scenarios and verification/validation completed. Fourth, it is formulated so that physical processes can be transparently improved through *modular code updates* as the science improves. Fifth, NAIRAS output can be readily *integrated with independent data streams* to improve the current epoch and near-term forecast radiation weather. Sixth, there is a large and *competent community* (NASA centers, research universities, and space weather industry) that collaborates to improve NAIRAS. Seventh, it has been to *represent radiation climatology well* on short time scales. These features lead to high confidence that NAIRAS will continue to evolve and form the foundation for a national U.S. capability to manage radiation risks to aviation. As other models evolve, we can imagine that ensemble modeling will also contribute a risk management pathway, but we do not address that capability here.

# 3 The Gannon Storm

## 3.1 Gannon storm nomenclature

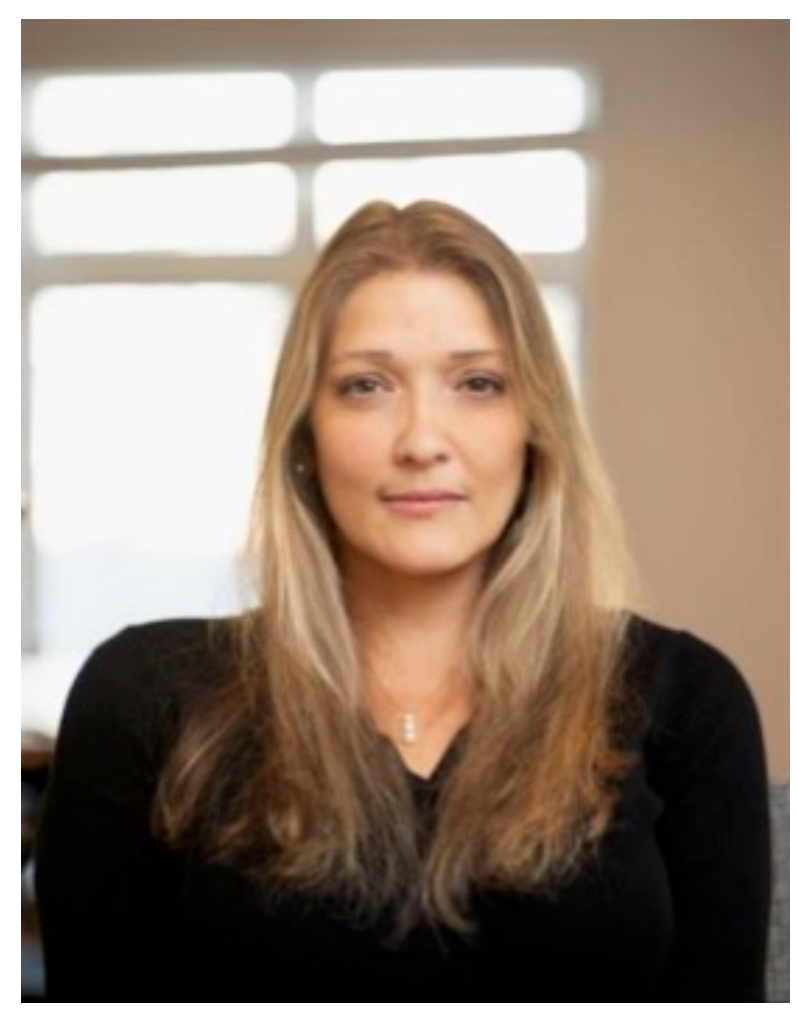

Fig. 3. Dr. Jenn Gannon, 2024.

The Gannon storm derives its name in recognition and memory by the broad space weather community of Dr. Jennifer Gannon. Dr. Gannon (Figure 3) was a leading international space weather physicist [Pulkkinen et al., 2024; Lugaz et al., 2024]. She passed away on May 2, 2024, just as the largest geomagnetic storm of solar cycle 25 began. Dr. Gannon was a key member of the SWAG and a significant contributor to its end user survey report described above. Dr. Gannon's scientific endeavors spanned radiation belt electron dynamics, geomagnetic storms, geomagnetically induced currents, and ground-based magnetic field disturbance instrumentation. Her extensive contributions covered fundamental scientific research, applied sciences and operational applications for the benefit of a range of end-users. She was also active in space physics and space weather leadership policy as editor of the American Geophysical Union's *Space Weather Journal* (SWJ) and the chair of the American Commercial Space Weather Association (ACSWA); both SWJ and ACSWA have helped shape the direction of the U.S. national space weather enterprise.

## 3.2 Long duration storm period

The Gannon storm started around May 2, 2024, when solar active regions (AR) 3663 and 3668 began appearing on the Earth-facing side of the Sun, having come over the solar East limb. The storm period (Figure 4) was marked by significant solar activity, including powerful solar flares, coronal mass ejections, solar energetic particle events, and a Forbush decrease.

Figure 4 provides a graphical timeline of the storm's evolution as it developed and receded between May 5 at 00 UT and May 17 at 24 UT. It includes 25 M5 class or larger flares where 17 of the flares were in the X-class category (top panel of Figure 4). There were 3 SEP events on May 9, May 11, and May 13 associated with X2.2, X5.8, and M6.6 flares, respectively. SEP #2 on May 11, 2024 was relevant to this study with a start time of approximately 01:40 UT, a peak time of approximately 02:45, and a long duration extending for the next two days. The GOES >10 MeV, >50 MeV, and >100 MeV proton measurements in the second panel show the very quick arrival time of SEP particles marked with the vertical magenta lines. The NOAA scale G4–G5 geomagnetic storm itself started around 15 UT on May 10 when a combination of flare-induced magnetic clouds, or CMEs, arrived at Earth. The third panel red vertical line marks the strong southward $B_z$ component of the interplanetary magnetic field (IMF). The fourth panel shows the solar wind speed, which rapidly changed from around 500 km $s^{-1}$ to over 700 km $s^{-1}$ at the same time.

Neutron monitor data (Figure 10 below) identifies that the Forbush decrease started around 21 UT on May 10 as marked by the black vertical line (Figure 4). The low levels of GCRs continued well into May 12, i.e., beyond the extent of flight UAL 990 and beyond the onset of SEP #2. As the GCRs are the main contributor to the dose rates, the lowered dose rates from the Forbush decrease persisted for the entire flight, i.e., beyond the onset of the small and transient SEP#2. This was the time during the initial G4 storm that the CMEs' magnetic field acted as a barrier to reduce the influx of GCRs into Earth's atmosphere. It effectively shielded the planet from lower-energy GCR particles. The G5 main storm event occurred within 3 hours of the Forbush decrease. Finally, the Figure 4 green vertical lines show the flight duration for United Air Lines (UAL) 990 between San Francisco and Paris on May 10 – 11, 2024. This flight carried an ARMAS FM7 radiation measuring unit and it captured the dose during the flight through all the major events, as discussed below.

# 4 Two UAL 990 flights

## 4.1 ARMAS dosimetric definitions

The following dosimetric quantity definitions are used by ARMAS and are important for understanding radiation exposure in human tissue.

- Absorbed Dose in Silicon, D(Si), is the amount of energy absorbed by silicon per unit mass. This fundamental radiation measurement quantity has units of Gray (Gy). ARMAS reports this in one-millionth Gray (μGy) and one-millionth Gray per hour (μGy $h^{-1}$).
- Absorbed Dose in Tissue, D(Ti), is the amount of energy absorbed by human tissue per unit mass. This fundamental radiation measurement quantity has units of Gray (Gy). ARMAS reports this in one-millionth Gray (μGy) and Absorbed Dose Rate in Tissue, dD(Ti)/dt, in one-millionth Gray per hour (μGy $h^{-1}$).
- Average Quality Factor, Q, scales the exposure in a specific radiation field to the potential biological risk. This calculated value is not reported by the ARMAS app, but its calculation can be found in ARMAS publications [Tobiska *et al.,* 2018] and ARMAS metadata records.
- Dose Equivalent, H, is the radiation quantity used to report a person's exposure for regulatory, medical, and scientific purposes. Regulatory limits are expressed in units of Dose Equivalent. Dose Equivalent is calculated by multiplying the Absorbed Dose (D) in tissue times the Average Quality Factor (Q) ($H = D \times Q$). Units for Dose Equivalent are reported in Sieverts (Sv). ARMAS reports this in one-millionth Sievert (μSv) and Dose Rate Equivalent, dH/dt, in one-millionth Sievert per hour (μSv $h^{-1}$).

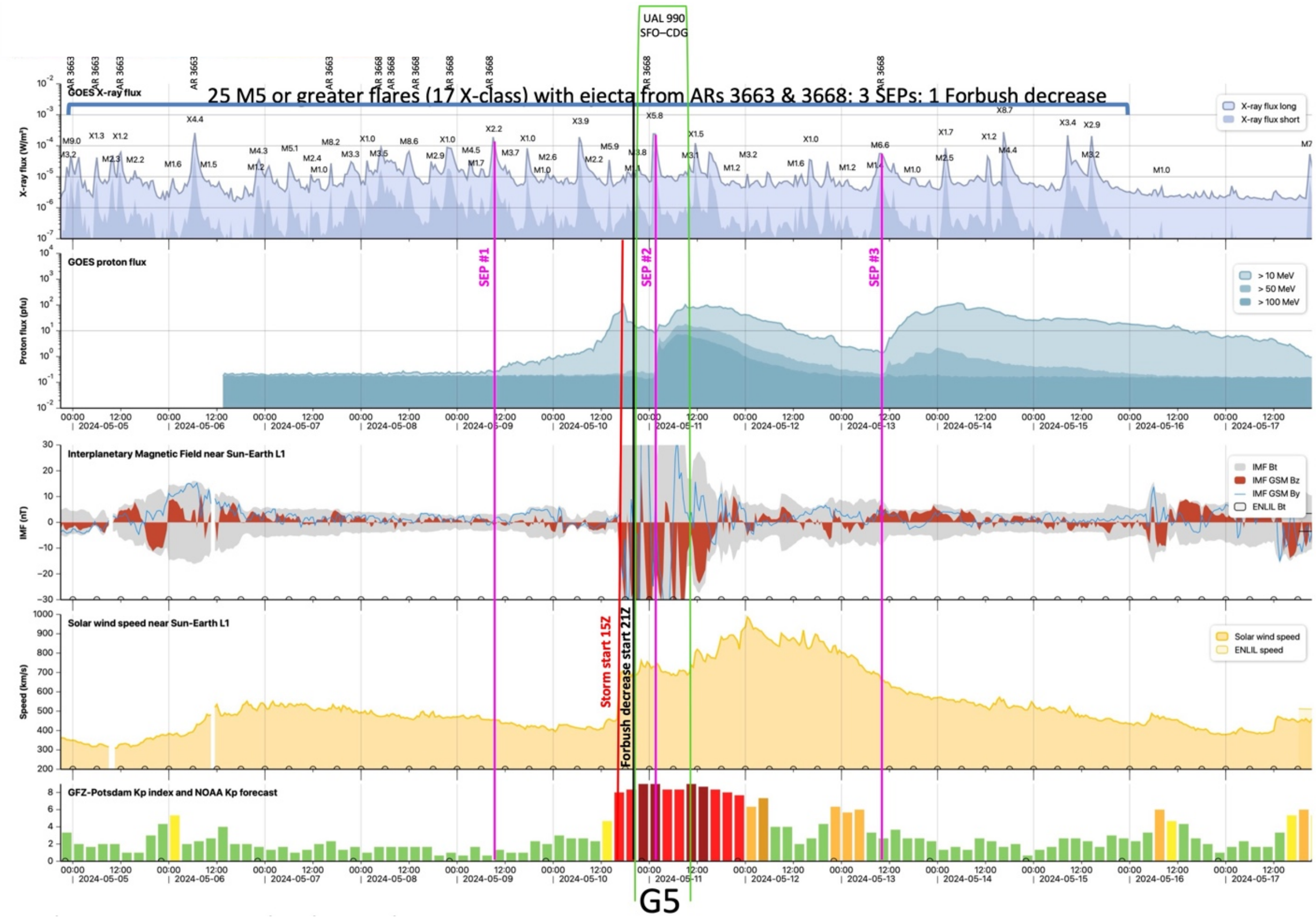

Fig. 4. Flare, CME, SEP, and Forbush decrease phenomena of the Gannon storm between May 5 – 17, 2024. The top panel identifies the GOES-16 0.1–0.8 nm X-rays during the time frame with class of flare labeled and the active region (AR) association also identified at the top of the panel. The second panel shows the GOES proton flux for >10, >50, and >100 MeV. SEP events are marked by magenta vertical lines. The third panel provides the IMF in units of nT where red indicates southward Bz. The start of the Gannon G4-G5 storm is marked with a red vertical line. The fourth panel provides the solar wind speed from DSCOVR and/or ACE. The fifth panel uses histogram bars to mark the Kp index. The UAL 990 flight start, and end times are demarcated by the vertical green lines while the start of the Forbush decrease is identified by a black vertical line. [Courtesy of KNMI]

- Ambient Dose Equivalent, H*(10), is a quantity developed for operational field measurements. It reports the average absorbed dose from all radiation at a depth of 10 mm inside a tissue equivalent material such as a human torso phantom. Units for Ambient Dose Equivalent are reported in Sieverts (Sv). ARMAS reports this in one-millionth Sievert (μSv) and Ambient Dose Rate Equivalent, dH*(10)/dt, in one-millionth Sievert per hour (μSv $h^{-1}$).
- Effective Dose, E, is not measurable but is derived using a mathematical system that weights the Dose Equivalent received by each separate organ tissue (T) in the human body, $H_T$, by a unique weighting factor ($W_T$). This weighting factor considers the specific sensitivity of each organ to different types of radiation. When the product of these calculations for each organ is summed, the total value is the Effective Dose, E. Calculating the Effective Dose is especially useful in determining radiation risk for individuals that have received partial body irradiations. For individuals receiving uniform full body irradiations (non-localized partial body irradiations), the risk calculated by Effective Dose is the same as that measured in Dose Equivalent. Units for Effective Dose are reported in Sieverts (Sv). ARMAS reports this in one-millionth Sievert (μSv) and Effective Dose Rate, dE/dt, in one-millionth Sievert per hour (μSv $h^{-1}$).
- Dose index, D. The D-index was developed to provide warnings of elevated radiation levels. As used by ARMAS, it is based on the radiation exposure from solar particle and radiation belt precipitation summed with background galactic cosmic rays. It is created from the effective dose rate, which can be derived from either measurements or models. The D-index range from D0 to D8 covers a wide range of radiation exposures at aviation altitudes. D0, D1 and D2 levels are for quiet space weather conditions. D3 is for elevated exposure from more particles coming into the atmosphere and can be used by air traffic management to trigger a radiation warning. D4 and higher indicate radiation alerts; they occur infrequently but during large solar events.

### 4.2 UAL 990 background environment for 2024 and 2025 flights

**Table 1. ARMAS v11.39 dosimetric data collected during the UAL 990 2024 & 2025 flights**

| ***Quantity*** | ***2024 Value*** | ***2025 Value*** | ***Units*** |
|---|---|---|---|
| Flight total measured absorbed dose, D(Si) | 13.16 | 14.28 | µGy |
| Flight total derived absorbed dose, D(Ti) | 19.45 | 21.41 | µGy |
| Maximum flight derived absorbed dose rate in silicon, dD(Si)/dt | 4.20 | 4.20 | µGy $h^{-1}$ |
| Median flight derived absorbed dose rate in tissue, dD(Ti)/dt | 2.48 | 2.52 | µGy $h^{-1}$ |
| 1 standard deviation (1$\sigma$) flight derived absorbed dose rate, dD(Ti)/dt | 0.97 | 1.16 | µGy $h^{-1}$ |
| Flight derived total dose equivalent, H | 41.60 | 46.55 | µSv |
| Median flight derived dose equivalent rate, dH/dt | 5.31 | 5.48 | µSv $h^{-1}$ |
| 1σ flight derived dose equivalent rate, dH/dt | 2.07 | 2.52 | µSv $h^{-1}$ |
| Flight derived total ambient dose equivalent, H*(10) | 64.02 | 71.65 | µSv |
| Median flight derived ambient dose equivalent rate, dH10/dt | 8.17 | 8.43 | µSv $h^{-1}$ |
| 1σ flight derived ambient dose equivalent rate, dH10/dt | 3.19 | 3.89 | µSv $h^{-1}$ |
| Flight mean total effective dose, E | 78.75 | 89.61 | µSv |
| NAIRAS v3 modeled flight mean total effective dose, E | 91.92 | 80.83 | µSv |
| Median flight derived effective dose rate, dE/dt | 10.09 | 10.64 | µSv $h^{-1}$ |
| 1σ flight derived effective dose rate, dE/dt | 3.94 | 4.90 | µSv $h^{-1}$ |
| Flight estimated median Quality factor for range of cutoff rigidities | 2.14 | 2.17 | — |
| Flight estimated 1σ Quality factor for range of cutoff rigidities | 0.02 | 0.01 | — |

### 4.3 UAL 990 flight details on May 10–11, 2024 during the Gannon storm

UAL 990's B777-200 aircraft (Figure 5) departed San Francisco (SFO) on May 10, 2024, at 21:40 UT, bound for Paris (CDG) on a 11.25-hour duration flight. An ARMAS FM7 was flown on board. Excellent data were recorded for the entire flight. Of the 676 1-minute data records, science quality data comprised 98.52% of records.

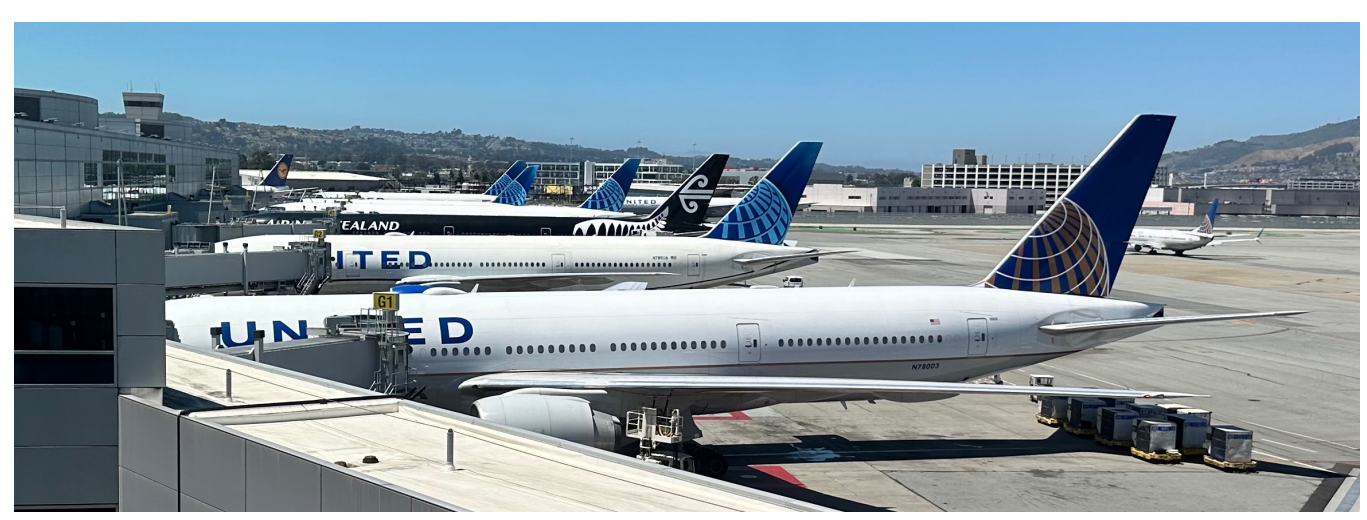

Fig. 5. UAL 990 B777 in foreground at SFO Gate G1 on May 10, 2024, before flight to CDG.

Prior to the departure, starting on May 9, 2024, NOAA's Space Weather Prediction Center (SWPC) proactively advised the aviation sector of possible large geomagnetic storms in the coming days. On May 10, 2024, FAA Air Traffic Control issued an advisory to all carriers on the developing storm conditions and the potential for communication outages at higher latitudes and GPS navigation outages or degradations.

This flight's planned route is normally a great circle from San Francisco, over the Hudson Bay, across Greenland, over Iceland, above the UK, and to Paris was deviated prior to flight by UAL operations because of the major geomagnetic storm predictions by SWPC. Typically, the aircraft flies at 11.6 km (38,000 ft.) to 12.2 km (40,000 ft.) in altitude and will reach latitudes approaching

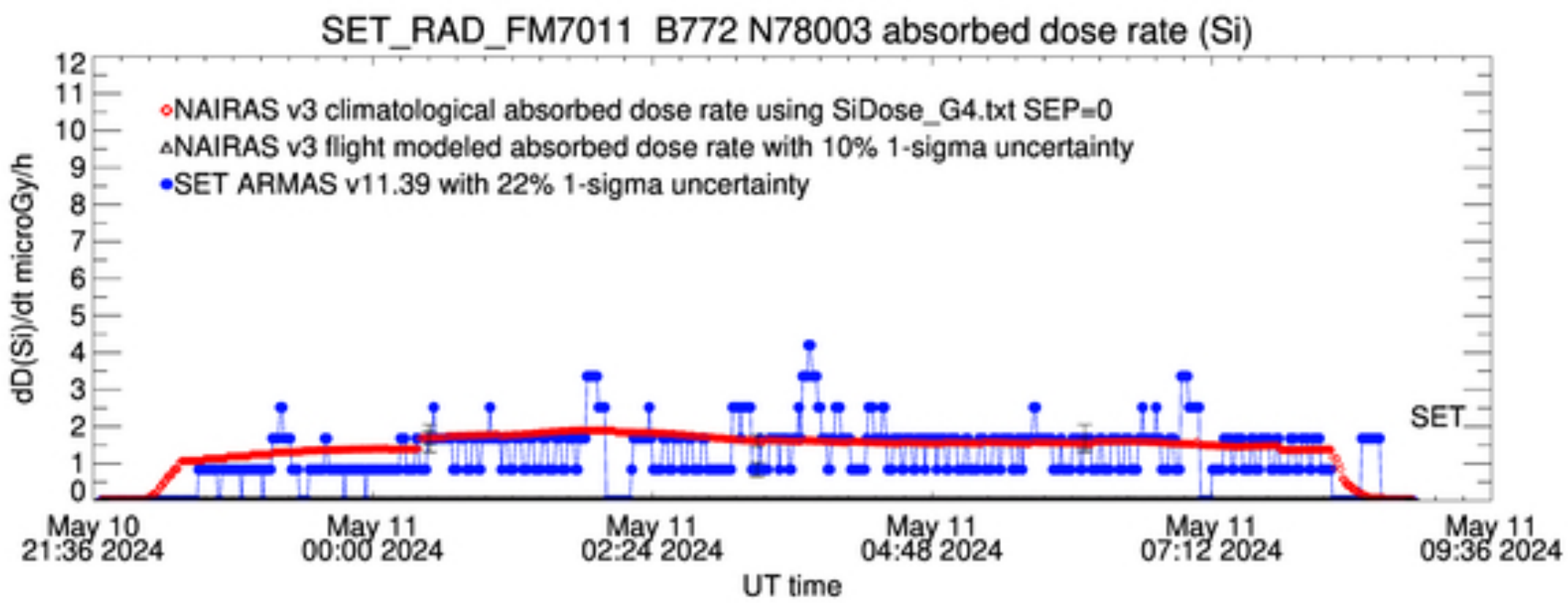

Fig. 6. Absorbed dose rate in silicon for the UAL 990 flight SFO to CDG on May 10–11, 2024. The Y-axis shows the absorbed dose rate in silicon in microGy/hour. The legend identifies the NAIRAS v3 climatology data for G4 conditions (red diamond), NAIRAS v3 flight modeled data with 10% uncertainty (black triangle – not calculated for this flight), and ARMAS v11.39 measurements (blue dot) with 22% uncertainty. There was no SEP file used at the start of the flight and SEP#2 extended for much of the flight after 01:40 UT.

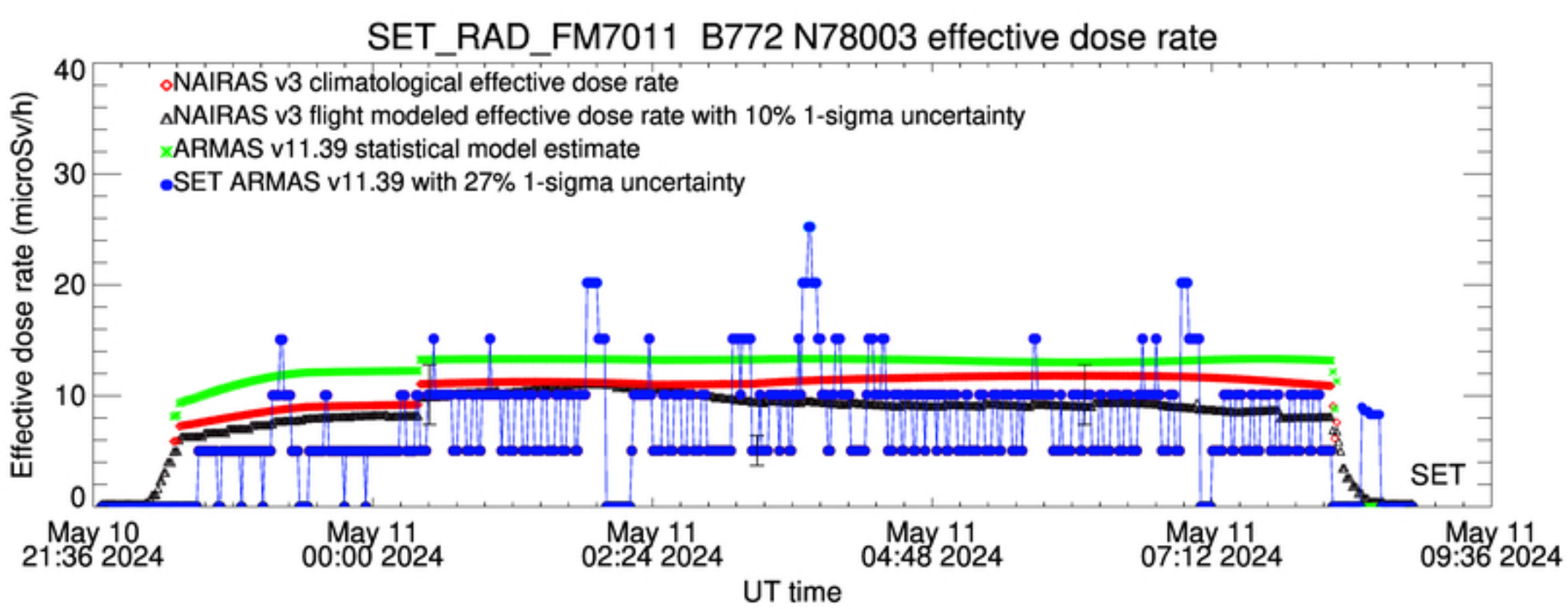

Fig. 7. Effective dose rate for the UAL 990 flight SFO to CDG on May 10–11, 2024. The Y-axis shows the effective dose rate in microSv/hour. The legend identifies the NAIRAS v3 climatology data (red diamond), NAIRAS v3 flight modeled data with 10% uncertainty (black triangle), ARMAS v11.39 statistical model estimates (green asterisk), and ARMAS v11.39 measurements (blue dot) with 27% uncertainty.

65 degrees north. The pre-flight deviated path was selected due to communication loss risks for high-latitude transatlantic routes between western Europe and eastern coast of North America. In this case, instead of flying the great circle route, UAL 990 flew at 40 to 43 degrees north latitude across the continental United States, south of Nova Scotia, over the Atlantic Ocean, and into Paris (10116 km). The flight altitudes ranged from about 10.4 km (34,000 ft.) to 11 km (36,000 ft.) and the maximum latitude reached was 51 degrees north off the coast of France. The total flight effective dose was 79 μSv, less than the total dose during higher latitude, quiet geomagnetic conditions.

The environment during this flight included: *i)* an extreme space weather event; *ii)* average Kp of 8; *iii)* average Ap of 295; *iv)* average NOAA G level of G4; *v)* one SEP event (SEP #2 in Figure 4); *vi)* cruise altitude of 9875.52 m defined as 0.90 of the maximum altitude; *vii)* median altitude of 10972.80 m; *viii)* median altitude standard deviation of 273.11 m; and *ix)* median cutoff rigidity of 0.56 GV. The ARMAS v11.39 dosimetric data collected during the flight is shown in Table 1.

4.4

## 4.4 UAL 990 May 2024 flight radiation measurements

The flight profile during radiation data collection is shown in Figures 6 and 7. Figure 6 is the measured dose rate in silicon and Figure 7 is the derived effective dose rate. Using Figure 7 as an example, several features are seen: *i)* the flight takeoff (left side) and landing (right side) have dose rates of "0" while on the ground – this is normal; *ii)* as the flight ascends above 8 km (altitudes not shown) the dose rate rises from "0" to some value; *iii)* ARMAS derived effective dose rates (Figure 7) show distinct variability above a baseline threshold of approximately 5 μSv $h^{-1}$ and reach a maximum value of 25 μSv $h^{-1}$ at 03:44 UT, following the peak time of the SEP #2 event; *iv)* the range of other ARMAS values is between 10 and 15 μSv $h^{-1}$; *v)* the ARMAS data resolution appears digitized due to instrument channel reporting thresholds for accumulated dose – this is normal; *vi)* in addition to the blue ARMAS connected dots with 27% uncertainty there are three other symbols – NAIRAS v3 climatological values for G4 conditions (red diamond), NAIRAS v3 flight modeled values with 10% uncertainty at the time of the flight (black triangle) and, ARMAS statistical model values (green asterisk); *vii)* ARMAS statistical values are slightly higher than NAIRAS v3 climatology or flight modeled data although they all represent the approximate GCR background for the given altitude, latitude, and geomagnetic conditions; and *viii)* ARMAS measured values above the GCR background are likely due to non-SEP but precipitated VAB electrons and their energy deposition processes.

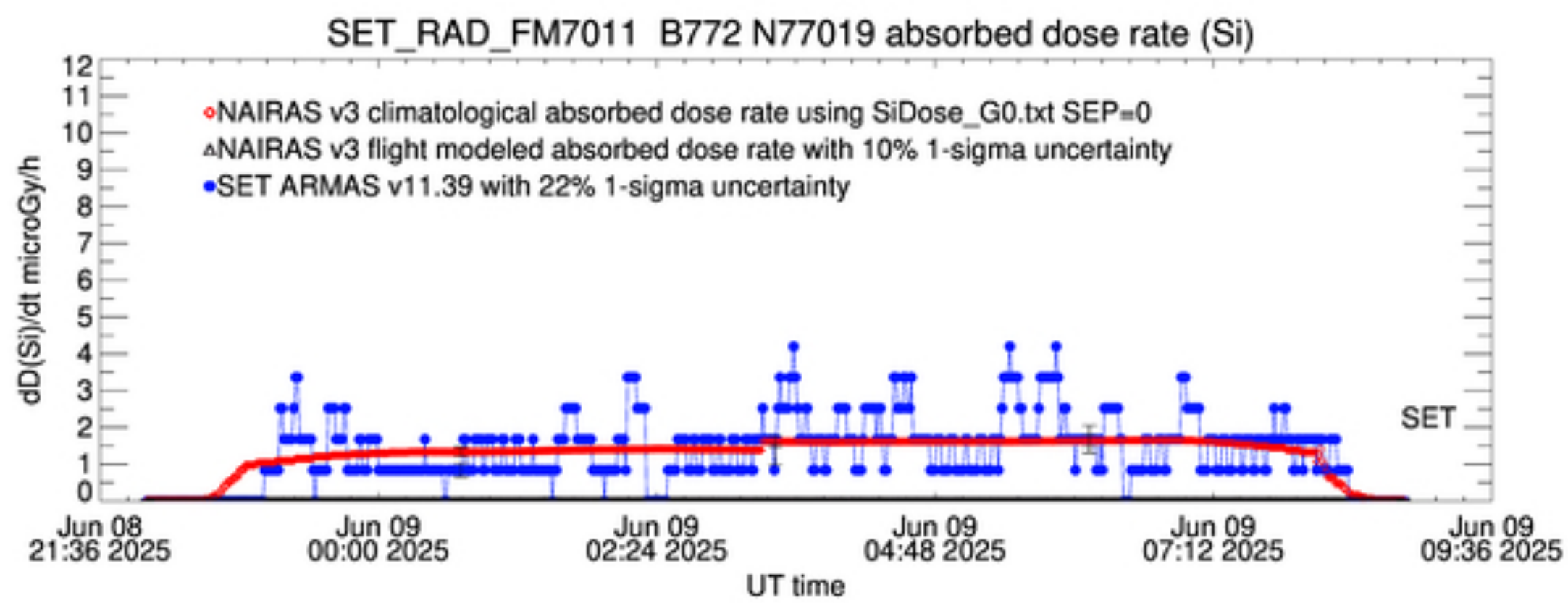

Fig. 8. Absorbed dose rate in silicon for the UAL 990 flight SFO to CDG on June 8–9, 2025. The Y-axis shows the absorbed dose rate in silicon in microGy/hour. The legend identifies the NAIRAS v3 climatology data for G0 conditions (red diamond), NAIRAS v3 flight modeled data with 10% uncertainty (black triangle – not calculated for this flight), and ARMAS v11.39 measurements (blue dot) with 22% uncertainty. There was no SEP file used at the start of the flight.

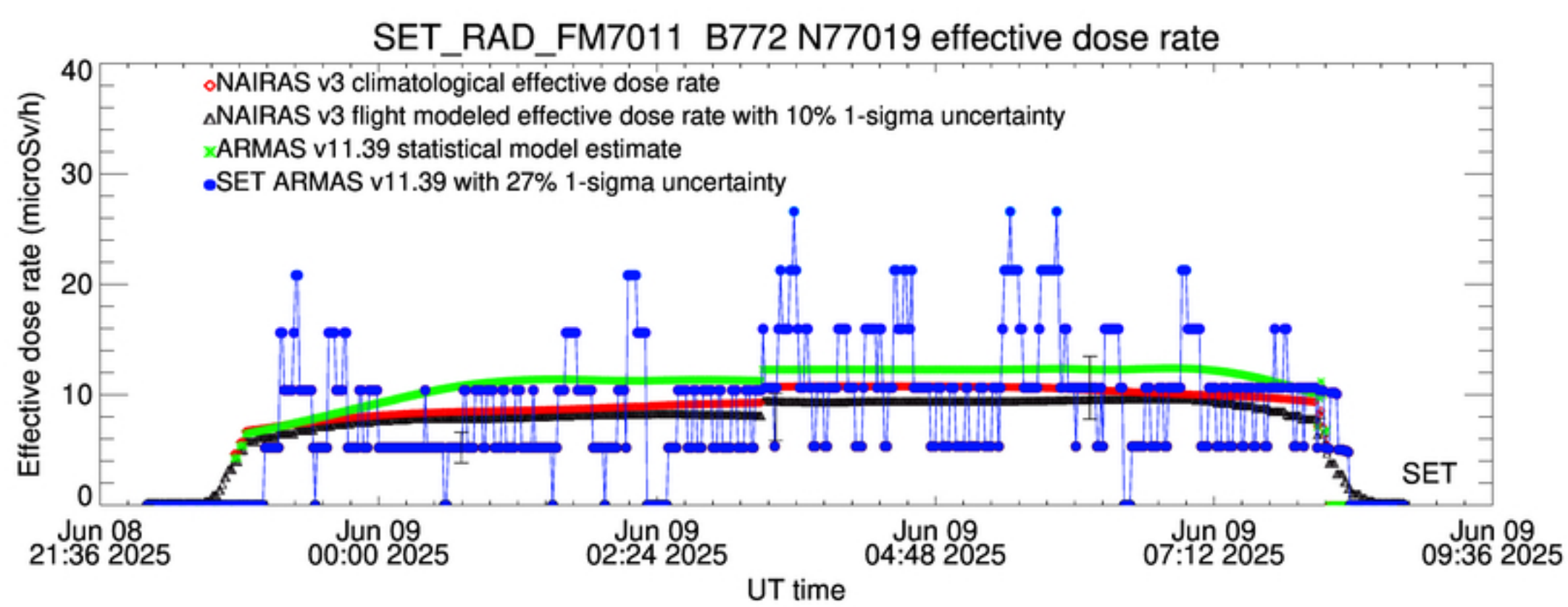

Fig. 9. Effective dose rate for the UAL 990 flight SFO to CDG on June 8–9, 2025. The Y-axis shows the effective dose rate in microSv/hour. The legend identifies the NAIRAS v3 climatology data (red diamond), NAIRAS v3 flight modeled data with 10% uncertainty (black triangle), ARMAS v11.39 statistical model estimates (green asterisk), and ARMAS v11.39 measurements (blue dot) with 27% uncertainty.

During the G5 period of the Gannon Storm, the flight was between Chicago and Nova Scotia. The GCR background, as measured with the ARMAS, was between 5–10 μSv $h^{-1}$. The effective dosage for the entire flight was 79 μSv. This was less than would have been expected during a higher latitude flight. Because United 990 was flying a longer distance at lower altitudes and latitudes, one would have expected higher fuel consumption and a longer flight time (~12 hours vs. 10.83 hours for the typical flight). However, due to strong tailwinds over the Atlantic Ocean, the United 990 flight took only a half hour longer than originally planned.

### 4.5 UAL 990 flight details on June 8–9, 2025 during quiet conditions

As a comparison to the May 10–11, 2024 UAL 990 flight, a second flight was conducted under non-storm, geomagnetically quiet conditions. An identical B777-200 aircraft designated UAL 990 departed San Franscisco (SFO) on June 8, 2025, again at 21:40 UT, bound for Paris (CDG) on a 10.83-hour duration flight. An ARMAS FM7 was flown on board with excellent data recorded for the entire duration. Of the 652 1-minute data records, the data was science quality for 100.0% of the records. The ARMAS v11.39 dosimetric data collected during the flight is shown in Table 1.

The flight took the great circular route from San Francisco over North Hudson Bay and into Paris (9117 km) at an altitude of 10.7 km (35,000 ft.) to 11.6 km (38,000 ft.), reaching a maximum of 63 degrees north latitude. The total flight effective dose was 90 μSv, i.e., more than during the G5 event at lower latitudes.

### 4.6 UAL 990 June 2025 flight background environment

The background environment during this flight included: *i)* quiet space weather conditions; *ii)* average Kp of 3; *iii)* average Ap of 15; *iv)* average NOAA G level of G0; *v)* no SEPs (0); *vi)* cruise altitude of 10424.16 m defined as 0.90 of the maximum altitude; *vii)* median altitude of 11582.40 m; *viii)* median altitude standard deviation of 418.99 m; and *ix)* median cutoff rigidity of 0.013 GV. The dosimetric data collected the flight is summarized in Table 2.

### 4.7 UAL 990 June 2025 flight radiation measurements

The flight profile during radiation data collection is shown in Figures 8 and 9. Figure 8 is the measured dose rate in silicon and Figure 9 is the derived effective dose rate. Using Figure 9 as an example, several features are seen: *i)* the flight takeoff and landing are similar to the May 2024 flight; *ii)* ARMAS derived effective dose rates (Figure 9) show distinct variability above a baseline threshold of approximately 5 μSv $h^{-1}$ and reach a maximum value of 27 μSv $h^{-1}$ at 03:34, 5:26, and 5:50, UT; *iii)* the range of other ARMAS values is between 10 and 21 μSv $h^{-1}$; *iv)* the blue ARMAS connected dots are also complemented with the three types of symbols as in Figure 7; and *v)* ARMAS measured values above the GCR background are likely due to non-SEP but precipitated VAB electrons and their energy deposition processes, as in Figure 7.

During the quiet geomagnetic conditions, while the flight was at higher latitudes than during the Gannon storm, the GCR background, as measured with the ARMAS, was between 5–10 μSv $h^{-1}$. The total effective dose for the entire flight was 90 μSv. This is a typical value at this altitude, latitude range, and time duration. One might normally expect an entire accumulated dose equivalent to one chest X-ray of 100 μSv for this flight. This effective dose was greater than what was measured during the Gannon storm during a lower latitude flight for longer duration. We discuss these differences in the next section as they are related to the topic of validating dose mitigation strategies.

### 4.8 Radiation measurements summary

The net result for the May 2024 UAL 990 flight is that ARMAS obtained about 14% less total effective dose (79 µSv) during an extreme geomagnetic storm than was found on a similar flight (June 2025 UAL 990 flight), same aircraft, but during quiet geomagnetic conditions (90 µSv). Although the 2025 flight was at higher latitudes with less cutoff rigidity, $R_c$, the significant factor for the reduced total effective dose was due to 3 factors: Forbush decrease, lower altitude, and lower latitude.

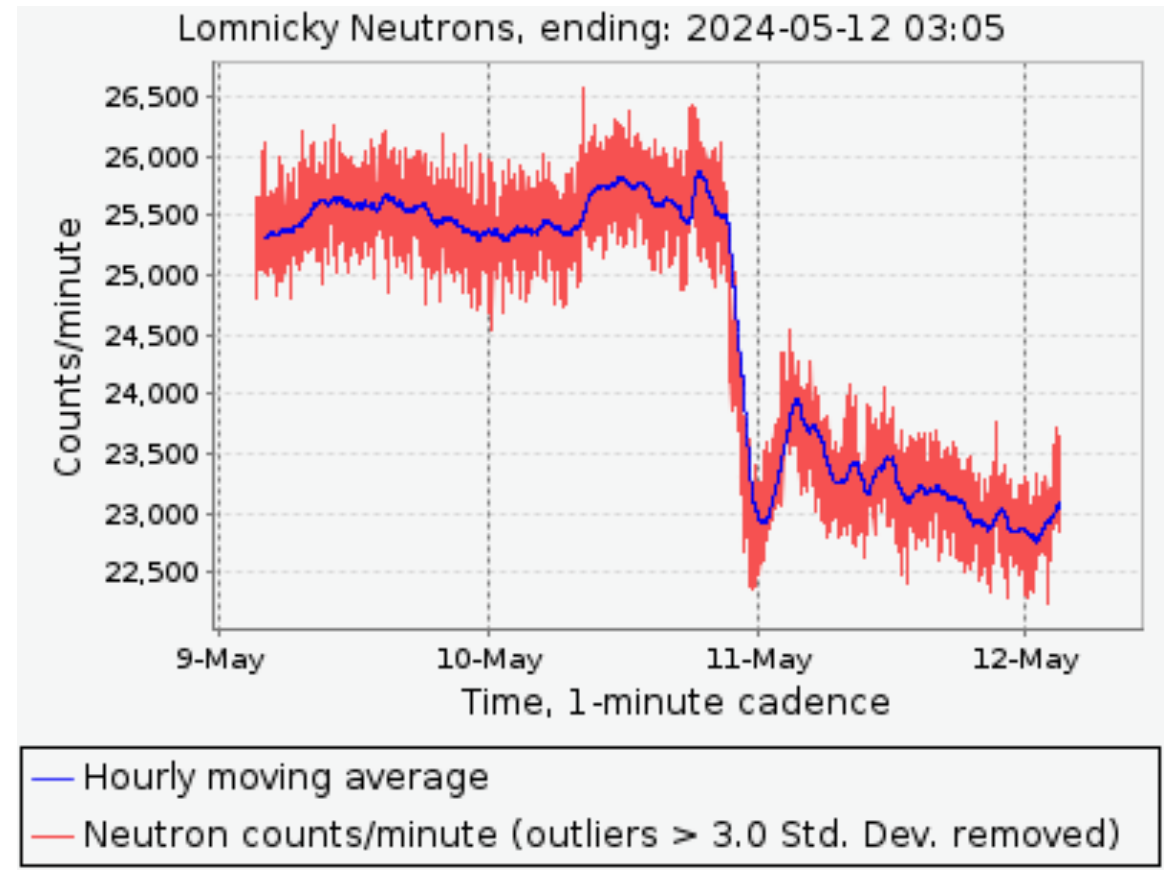

Fig. 10. Lomnicky neutron data during the Gannon storm Forbush decrease.

**Effect of Forbush decrease.** During the Gannon storm, a Forbush decrease started an hour before takeoff of UAL 990. The Lomnicky neutron data (Figure 10) saw an 11% decrease in GCRs between 21–24 UT on May 10, 2025. This phenomenon played a dual role: *i)* it demonstrated the strength of the magnetic cloud arriving at Earth and *ii)* it reduced the number of lower energy protons entering the Earth's atmosphere, which then linearly decreased the "floor" of GCR radiation exposure from fewer incoming particles. The entire planet saw a decreased background radiation environment during this period, which reduced the exposure hazard on the order of 10% estimated from that percentage of reduced particle flux. Subsequent SEP and VAB precipitation were then added to that background radiation "floor."

**Effect of altitude.** The May 2024 flight path during the Gannon storm took it to lower altitudes as it flew across the northern continental U.S., south of Nova Scotia, and across the Atlantic to France. The mean cruise altitude of 9.88 km was 0.54 km lower than a nominal flight mean altitude of 10.42 km as demonstrated by the June 2025 flight. From a heuristic perspective, an e-folding scale height for doubled radiation is about 2 km for mid-latitudes so, without rigorous radiation field modeling, the lower altitude would account for a measurable exponential reduction in mean dose rate during the flight, i.e., exp(-0.54/2) = 0.76, or a 24% atmospheric shielding reduction from the exponential increase in atmosphere density going to lower altitudes.

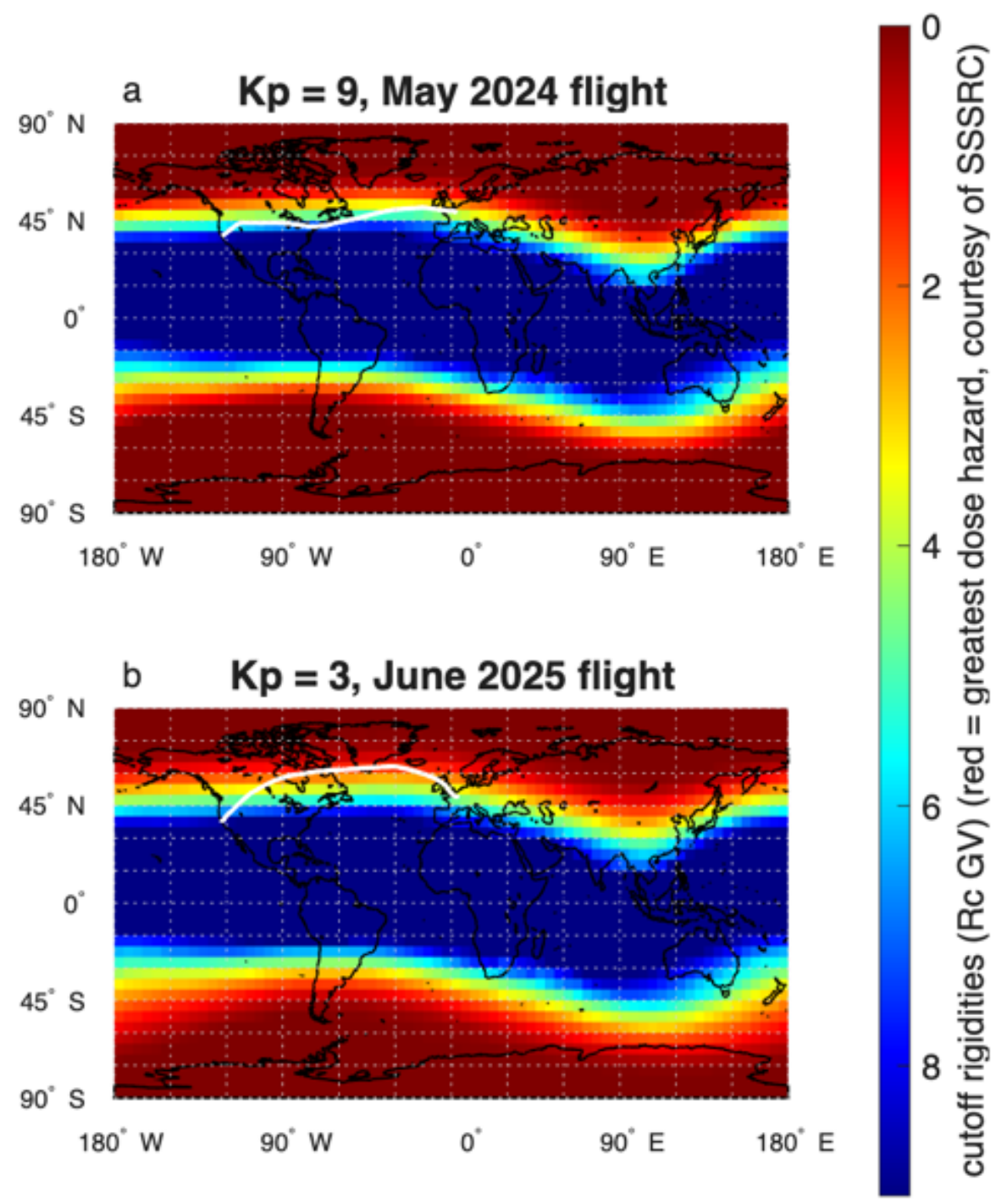

Fig. 11. (a) Cutoff rigidity map during the Kp = 9 timeframe of the Gannon storm in May 2024. The white line is the deviated flight path for UAL 990. (b) Cutoff rigidity map during the Kp = 39 timeframe of the quiet geomagnetic conditions in June 2025. The white line is the normal flight path for UAL 990. At 60° W longitude the cutoff rigidity, $R_c$, in 2024 was 5.2 GV and was 1.4 GV in 2025, showing the added magnetic field shielding in 2024 along the lower latitude flight path.

**Effect of latitude.** The May 2024 flight path during the Gannon storm took it to lower latitudes as it flew across the northern continental U.S., south of Nova Scotia, and across the Atlantic to France. The median cutoff rigidity, $R_c$, was 0.6 GV for this flight indicating that airspace in this route would have been exposed to the lower energy particles entering the northern hemisphere (Figure 11). Some relief with magnetic field shielding was accomplished along the lower latitude route although its significance was less than compared to atmospheric shielding. The percentage of radiation decrease due to lower latitudes is difficult to quantify here due to the unmodeled change in the cutoff rigidity from magnetic field structural changes. In Figure 11 we note that there is a sharp $R_c$ gradient across the north Atlantic in the sense that the magnetic field shielding scales linearly with precipitating particle energy. Heuristically, one might take an educated guess that the flight path through that gradient could equal a dose reduction comparable in magnitude to altitude. However, without rigorous modeling, that percentage cannot be further pinned down.

As shown in Figure 11, the May 2024 flight (11a) flew the higher $R_c$ route at lower latitudes compared to the June 2025 flight (11b), which flew the lower $R_c$ route and the higher latitudes. For comparative purposes, we note that the 60° W longitude location on each flight was half the way through the flights, was the time of the Gannon storm G5 event, and was near the maximum magnetic latitude for the quiet conditions. The May 2024 flight flew at $R_c$ = 5.2 GV at that longitude while the June 2025 flight flew at $R_c$ = 1.4 GV, i.e., less magnetic field shielding. We saw from measurements that the deviated route for May 2024 to lower latitudes and during an extreme geomagnetic storm contributed to a measurable total dose reduction for the entire flight compared to a normal flight at higher latitudes during quiet conditions in June 2025. In addition, a Forbush decrease also removed a population of lower energy GCR particles that, had they been present, would have increased the overall total dose.

# 5 Discussion

## 5.1 Radiation reduction

The deviated UAL 990 route on May 10–11, 2024 during the extreme Gannon storm to lower altitudes and lower latitudes had a quantifiable total effective dose reduction for the entire flight. A Forbush decrease removed a population of lower energy GCR particles that, had they been present, would have increased the overall total effective dose but the full extent of the dose reduction was not calculated in this study. The linear scaling effects of cutoff rigidity magnetic shielding of particles by energy played an important role, the reduced flux of GCRs from the Forbush decrease was linearly effective since less particles precipitated, and the decreased secondary particles due to lower flight altitude and exponentially increasing atmospheric shielding was significant.

The SEP #2 event that occurred during the May 2024 flight was also a factor and its extended duration contributed supplemental additional total dose, although the SEP contribution could not be uniquely separated from the GCR and possible VAB energy inputs. The timing of the start of the Forbush decrease occurred just prior to the start of the flight so the GCR reduction also could not be uniquely separated from other radiation sources. It is safe to say that conventional wisdom would expect more dose during an extreme storm (May 2024), SEP or no SEP, than during a quiet period (June 2025). Using this conventional wisdom, the total effective dose if UAL 990 in May 2024 had flown a great circle route could have been up to double the quiet period measured total dose. Calculations by the PANDOCA model [Schennetten et al., 2024] suggest, however, that a lower total dose of 14%–24% was more reasonable. No instrument was flown on that great circle route. The flight comparisons in this paper could not sort out these competing effects to obtain better than a qualitative assessment of a noticeable decrease in the expected effective dose. Flights at the same time on two aircraft in different locations would be the best possible experiment, for example.

A recent paper by Schennetten et al. [2024] explored the GCR Forbush decrease and SEP events of the Gannon storm on total dose using calculations of the PANDOCA model for a hypothetical great circle route of Frankfort (FRA) to Los Angeles (LAX) on May 10–11, 2024. They concluded that the addition of total dose to the flight due to the shift of cutoff rigidity equatorward was a minor contributor and consistent with the ARMAS measurements, although the latter could not distinguish the effects of latitude vs. cutoff rigidity separately. While the Schennetten et al. [2024] paper also explored the relative contributions of the Forbush decrease of dose and the SEP increase of dose, these combined effects leading to a 14%–24% total dose increase along the great circle route were not possible to separate out in the ARMAS dataset, especially when combined with latitude decrease. ARMAS did not measure pre-event Forbush decrease so that relative decline was not quantifiable in that flight. The paper also concluded that SEP dose rates were low for this SEP #2 event, a result that was supported by the ARMAS measurements.

The bigger story, however, is that this is the first time a radiation unit has been flown in such a large G5 storm at commercial aviation altitudes where CMEs, Forbush decrease, and SEP all occurred. The ARMAS results conclusively demonstrate an important policy success for the aviation community, i.e., that the radiation hazard from space weather can be measured as with ARMAS and shown to be mitigated using ALARA shielding strategies, i.e., two SWAG recommendations.

### 5.2 Mitigation strategy success

As mentioned above, the aviation sector, supported by scientific, radiation protection, policy, and regulatory communities, is moderately aware of space weather risks, particularly from radiation exposure. The depth of awareness varies across the aviation sector, but aviation operations and air traffic management do take actions to protect the industry from adverse space weather. In the past there has been planning for a major extreme event and that preparation paid off during the G5 Gannon Storm. The major United States carriers received early notifications that an event was possible and several of them acted to mitigate their risks. This includes aircraft operations at United Air Lines with respect to the May 10–11, 2024 flight 990 from SFO to CDG.

Space weather enterprise milestones that led to the documented success of the May 10–11, 2024 UAL 990 radiation exposure reduction by ARMAS include:

1) **1998–2025:** U.S. Government agencies and the national space weather enterprise (agencies, academia, industry) formulated and matured the *National Space Weather Program and Implementation Plan*;
2) **1999–2025:** NOAA SEL/SWPC hosted the annual Space Weather Workshop (SWW) with active participation from other agencies, academia, industry, and international stakeholders; in the Spring of 2005 UAL operations representatives introduced cross polar flights that were affected by space weather, adding carrier fuel/time/crew costs plus landing diversions as penalties for space weather;
3) **2000–2025:** FAA funded the development of the Civil Aerospace Medical Institute's (CAMI) climatological radiation model (up to CARI-7);
4) **2008–2025:** NASA Earth Science Division (ESD) and Heliophysics Division (HPD) funded the development of the NASA LaRC NAIRAS climatological radiation model;
5) **2011–2025:** NASA Small Business Innovation Research (SBIR), HPD, and Flight Opportunities (FO) funded Space Environment Technologies' (SET) commercial development and expansion of the ARMAS radiation detection real-time system;
6) **2013–2020:** The U.S. agencies' Space Weather Operations, Research, and Mitigation Subcommittee (SWORM) organized and proactively engaged the national space weather enterprise and Congress on the hazards of space weather to various sectors, including aviation;

7) **2013–2025:** U.S. advocated to ICAO for including space weather in its standards and recommended practices (SARPs); in 2018 ICAO published SARPs for space weather hazards to aviation in three areas: *i)* communications from high-frequency (HF) outages due to ionospheric disturbances from solar flares and geomagnetic storms; *ii)* navigation from Wide Area Augmentation System (WAAS)/GNSS outages due to ionospheric disturbances from solar flares and geomagnetic storms; *iii)* crew/passenger health from radiation exposure due to increased charged particle fluxes from at least GCRs and SEPs; *iv)* set up international centers for monitoring and advising aviation on space weather;
8) **2020:** PROSWIFT ACT became law with a mandate to the national space weather enterprise for developing mitigation activities for space weather risks to sectors including aviation;
9) **September 20–22, 2022:** NOAA SWPC held the first aviation testbed with stakeholders to refine responses to major space weather events;
10) **May 2–17, 2024:** a series of flare, coronal mass ejection (CME), and solar energetic particle (SEP) events in two separate active regions (ARs) began and lasted for a half solar rotation; one AR in each of northern and southern solar hemispheres; 25 flares of M5 class or larger (including 17 X-class flares); 3 separate SEP events; and 1 Forbush decrease occurred;
11) **May 7–15, 2024:** The Weather Company advised aviation customers of heightened space weather event awareness and preparedness for the next several days;
12) **May 9–15, 2024:** NOAA SWPC proactively advised the aviation sector of possible large geomagnetic storms in the coming days;
13) **May 10, 2024:** air traffic control (ATC) issued a NOTAM advisory to all carriers on the developing storm conditions; they advised of potential communication outages at higher latitudes in the North Atlantic (NAT) corridor; they advised of potential WAAS outages in CONUS;
14) **May 09–13, 2024:** multiple solar CMEs and SEPs combined to arrive at Earth within a small window of 5 days; 09 UT on May 9 the first SEP arrived; 15 UT on May 10 a G4 storm started; 21 UT on May 10 the first of two G5 events started, continuing into May 11; 21 UT on May 10 the Forbush decrease started; 02 UT on May 11 the second SEP arrived; 09 UT on May 13 the third SEP arrived; and
15) **May 10–11, 2024:** UAL 990 from SFO–CDG is preemptively deviated to a trans-CONUS and trans-Atlantic flight route for 11.2 hours; 15 UT on May 10 UAL operations advised crew that a deviation in waypoints to lower latitude and lower altitude was required to mitigate possible communication outages in North Atlantic (NAT) traffic routes; 20 UT on May 10 UAL gate personnel advised passengers that a deviation in waypoints was required to mitigate possible communication outages; 21 UT on May 10 ARMAS began radiation monitoring within the passenger cabin; 22 UT on May 10 aircraft took-off from SFO; 22-08 UT on May 10–11 GCR background was measured around 5–10 $\mu$Sv $h^{-1}$ by ARMAS; 02–03 UT on May 11 during SEP #2 ARMAS measured 25 $\mu$Sv $h^{-1}$; 08 UT on May 11 a total of 79 $\mu$Sv recorded by ARMAS.

**Findings:** This study found: *i)* lower GCR dose than expected partly due to Forbush decrease during entire flight; *ii)* measurement of SEP event in real-time and at FL360 (10 km); *iii)* SEP and non-GCR dose measurements during flight at night (possible source was X-rays and gamma rays resulting from radiation belt particle precipitation); *iv)* stakeholder awareness of space weather risks to aviation, including radiation, which had been heightened through years of preparation; *v)* event awareness and preparation was conducted successfully by multiple agency and commercial organizations before an extreme event arrived at Earth; *vi)* action was taken by major U.S. air carriers prior to flights that mitigated space weather risk from HF communication loss and from WAAS navigation outages although radiation risk was not a focus; *vii)* action taken by UAL operations to deviate UAL 990 to a lower altitude and lower latitude instead of great circle route resulted in noticeably lower total effective dose for the flight than would normally be expected;

and *viii)* other unintended results, including little loss in total flight time due to high tail winds of $200^+$ km $h^{-1}$ across the Atlantic.

# 6 Concluding remarks

## 6.1 First documented validation of aviation's radiation hazard mitigation strategy

Two UAL 990 flights using the B777-200 aircraft between San Francisco and Paris carried the same ARMAS FM7 radiation monitoring unit, where one flight occurred during the extreme geomagnetic storm (Gannon storm) May 10–11, 2024 and one flight occurred during quiet geomagnetic conditions June 8–9, 2025. The flights' results validated the strategy that multiple stakeholders in aviation radiation hazard mitigation have been pursuing for decades. That strategy is to use the ALARA shielding principle to reduce dose for activity that is under operational control during extreme space weather. One method is flying lower magnetic latitudes to gain more Earth magnetic field shielding and the other is flying lower altitudes to use atmosphere depth shielding. Both ALARA shielding methods are under carrier operational control and have now been validated with total dose measurements by ARMAS.

In addition to these two human controlled shielding methods, there was a third shielding process provided by nature in the G5 storm flight. A Forbush decrease at the beginning of the storm reduced the number of lower energy protons entering the Earth's atmosphere and that decreased the "floor" of GCR radiation exposure from cosmic rays. The entire planet saw decreased background radiation during this period, and this reduced the exposure hazard.

## 6.2 Data results summarized

The net result for the May 2024 UAL 990 lower latitude/lower altitude/longer distance flight was that ARMAS obtained about 14% less total effective dose (79 μSv) during an extreme geomagnetic storm than was found on a higher latitude/higher altitude/shorter distance flight (June 2025 UAL 990 flight), same aircraft, but during quiet geomagnetic conditions (90 μSv). Some variables in this comparison were the same: *i)* same aircraft type (B777-200); *ii)* same seating location inside the aircraft (business class); *iii)* same range for flight times (22–08 UT); *iv)* same ARMAS FM7011 instrument; and *v)* same ARMAS algorithm processing version (11.39).

Other variables were different: *i)* the 2024 flight was during a G5 extreme geomagnetic storm and a SEP event, which would have increased the dose, while the 2025 flight was geomagnetically quiet; *ii)* the 2024 flight had a 11% Forbush decrease, which would have reduced the GCR component of the 2024 planetary radiation at aviation altitudes, while the 2025 flight had no such decrease; *iii)* the 2024 flight mean cruise altitude was a half km lower (9.88 km), which would have provided more atmospheric shielding and reduced the radiation at aviation altitudes, compared to the 2025 flight (10.42 km); *iv)* the 2024 flight had a lower latitude flight path (51° N maximum), which increased the cutoff rigidity of $R_c$ = 5.2 GV, reduced the number of incoming particles, and reduced the radiation at aviation altitudes, while the 2025 flight had a higher latitude flight path (63° N maximum) with $R_c$ = 1.4 GV; and *v)* the duration of the 2024 flight was about 25 minutes longer (11.25 h) than the 2025 flight (10.83 h), which may have contributed an insignificant amount of additional total dose.

For the May 10–11, 2024 Gannon storm UAL 990 flight the net effect of the increased particles during the G5 event, the decreased particles from the Forbush decrease, the decreased particles from the lower altitude, the decreased particles from the lower latitude, and the increased particles from the longer duration resulted in an overall total effective dose lower than might be expected, as evidenced by the quiet period baseline June 8–9, 2025 UAL 990 flight. The natural shielding from the Forbush decrease and the human induced (airline operations) shielding using the denser atmosphere and the stronger magnetic rigidity were the dominant ALARA application successes in the May 2024 UAL 990 flight. This aviation stakeholder strategy was validated by this comparison.

The results of NAIRAS model simulations for the ARMAS data are presented in figures 6, 7, 8, and 9 along with the ARMAS measurements. SET's implementation of operational NAIRAS v3 does provide, in general, the GCR and SEP components of the radiation field but to extract out the specific components will require a more detailed study with the help of the NAIRAS team. A detailed SEP simulation by NAIRAS v3 operational code on SET was not available at the time of this paper.

### 6.3 Recommendations for future aviation radiation mitigation

This study validates aviation's radiation hazard mitigation strategy articulated by numerous stakeholders, which is to deviate flight paths to lower altitudes and lower magnetic latitude routes during major geomagnetic storms. This is the aviation industry's approach to mitigating HF communication outages and reducing risks from increased GNSS errors for take-off and landing navigation. Thus, lower latitude magnetic field shielding is an important risk reduction factor for radiation, communication, and navigation while lower altitude atmospheric shielding significantly reduces radiation risks.

## Data availability statement

The datasets presented in this study can be found in online repositories. The most recent data are available at the https://spacewx.com website under the Mission/Radiation Decision Aids/ARMAS Main Page/Data Archive links. The data at this archive can be accessed freely and publicly. The content of the archive includes both ascii text (data) files with a listing of the dosimetric parameters and space weather conditions in the minute-by-minute time records. There are also multiple figures for each flight showing dose rates by time, L-shell, and altitude.

## Author contributions

WKT: Conceptualization, Data curation, Formal Analysis, Investigation, Methodology, Project administration, Supervision, Writing – original draft, Writing – review and editing. BH: Methodology, Visualization, Data curation, Writing – review and editing. LD: Instrument development. KJ: Instrument firmware. JB: Writing – review and editing. KD: Data curation, Software. KW: Data curation, Software, App development. AS: Data curation, Software, App development.

## Funding

The authors declare that financial support was received for the research and/or publication of this article. Subcontract 09650000102835 from UCLA (PI Jacob Bortnik) to SET under NASA grant 80NSSC21K1308 provided financial assistance.

## Acknowledgements

The authors thankfully acknowledge NASA's Heliophysics Division, Flight Opportunities program, and Small Business Innovative Research (SBIR) program for extensive competitive funding assistance with ARMAS development and deployment over the past solar cycle.

# Conflict of interest

The authors declare that the research was conducted in the absence of any commercial or financial relationships that could be construed as a potential conflict of interest.

# Generative AI statement

The author(s) declare that no Generative AI was used in the writing of this manuscript.

# Publisher's note

All claims expressed in this article are solely those of the authors and do not necessarily represent those of their affiliated organizations, or those of the publisher, the editors and the reviewers. Any product that may be evaluated in this article, or claim that may be made by its manufacturer, is not guaranteed or endorsed by the publisher.

# Supplementary material

The Supplementary Material for this article can be found online at the location listed under Data Availability above.